# Point estimation for adaptive trial designs I: a methodological review

**David S. Robertson[1*], Babak Choodari-Oskooei[2], Munya Dimairo[3], Laura Flight[3], Philip Pallmann[4], Thomas Jaki[1,5]**

[1] *MRC Biostatistics Unit, University of Cambridge*
[2] *MRC Clinical Trials Unit at UCL*
[3] *School of Health and Related Research (ScHARR), University of Sheffield*
[4] *Centre for Trials Research, Cardiff University*
[5] *University of Regensburg, Germany*
[*] *Corresponding author: david.robertson@mrc-bsu.cam.ac.uk*

## Abstract

Recent FDA guidance on adaptive clinical trial designs defines bias as "a systematic tendency for the estimate of treatment effect to deviate from its true value", and states that it is desirable to obtain and report estimates of treatment effects that reduce or remove this bias. The conventional end-of-trial point estimates of the treatment effects are prone to bias in many adaptive designs, because they do not take into account the potential and realised trial adaptations. While much of the methodological developments on adaptive designs have tended to focus on control of type I error rates and power considerations, in contrast the question of biased estimation has received relatively less attention. This paper is the first in a two-part series that studies the issue of potential bias in point estimation for adaptive trials. Part I provides a comprehensive review of the methods to remove or reduce the potential bias in point estimation of treatment effects for adaptive designs, while part II illustrates how to implement these in practice and proposes a set of guidelines for trial statisticians. The methods reviewed in this paper can be broadly classified into unbiased and bias-reduced estimation, and we also provide a classification of estimators by the type of adaptive design. We compare the proposed methods, highlight available software and code, and discuss potential methodological gaps in the literature.

# 1. Introduction

Adaptive clinical trials allow for pre-planned opportunities to alter the course of the trial on the basis of accruing information[1–3]. This may include changes such as increasing the recruitment target (in sample size re-estimation designs), selecting the most promising treatment arms (in multi-arm multi-stage designs) or patient subpopulations (in population enrichment designs), shifting the randomisation ratio towards more promising arms (in adaptive randomisation designs), or terminating recruitment early for clear evidence of benefit or lack thereof (in group sequential designs)[4]. Despite some additional complexities when implementing such trials[5], they are increasingly being used in practice due to their attractive features, adding flexibility to a trial design whilst maintaining scientific rigour[6–8]. The challenges of the COVID-19 pandemic have also recently accelerated their use[9,10].

To date most of the research undertaken has focused on the design of such studies and the associated question of maintaining desirable operating characteristics related to hypothesis testing (namely type I error and power) rather than estimation. General methods on the basis of $p$-value combination[11,12] and conditional error functions[13,14] have been proposed that are applicable to a wide range of adaptive designs as well as specialised methods for specific designs such as multi-arm multi-stage (MAMS) designs[15,16], adaptive enrichment designs[17,18], response-adaptive randomisation (RAR) designs[19,20], and sample size re-estimation[21,22] (see Section 1.1 for definitions of the different adaptive designs considered in this paper).

In contrast, the question of estimation of treatment effects in an adaptive clinical trial has received comparatively less attention, as reflected in the recent FDA guidance on adaptive designs[23(p8)] which states "Biased estimation in adaptive design is currently a less well-studied phenomenon than Type I error probability inflation". This paper is the first in a two-part series that studies the issue of potential bias in point estimation for adaptive designs. In the current paper (part I), we review methods for unbiased and bias-reduced estimation of treatment effects after an adaptive clinical trial and critically discuss different approaches. In part II, we consider point estimation for adaptive designs from a practical perspective, including a set of guidelines for best practice.

While equally important, the construction of related quantities for inference, such as confidence intervals or regions, is beyond the scope of this series so we signpost the interested reader to related literature[24,25]. Moreover, our focus is on bias in the statistical sense, and we do not consider issues such as operational bias in adaptive designs[2,3,26] or the impact of publication bias[27]; see also recent reference articles[28,29] for discussion of other types of biases in randomised controlled trials.

The structure of the paper is as follows. After defining different types of adaptive designs in Section 1.1, we introduce key concepts and the different definitions of bias in Section 2. We describe the search strategy used for a systematic review of the methodological literature in Section 3. We present a summary of the results of the review grouped by the type of adaptive design in Section 4. Sections 5 and 6 then provide detailed descriptions and explanations of the different types of point estimators, divided into methods for *unbiased estimation* (which aims to completely remove bias) in Section 5 and *bias-reduced estimation* (which aims to reduce the magnitude of the bias, but not necessarily to eliminate it) in Section 6. In the supplementary information, we provide an annotated bibliography giving further details of all the papers included in the systematic review. The overall conclusions are discussed in Section 7.

## 1.1 Glossary of adaptive designs

Table 1 gives definitions of the different types of adaptive designs considered in this paper, including the different terminologies used for them. These definitions are based on those provided by Pallmann et al.[1] and Burnett et al.[4], who note that there can be ambiguous terminology used in the literature. Adaptive designs can combine multiple adaptive features within a single trial, see Dimairo et al.[2] for examples.

| **Design** | **a.k.a.** | **Definition** |
|---|---|---|
| Group sequential | Multi-stage design | Allows for early stopping of the trial for safety, futility or efficacy.<br><br>Group sequential designs can either be two-arm or single-arm trials, with Simon's two-stage design being a common example of the latter. For the generalisation to multiple treatment arms, see 'Multi-arm multi-stage designs'.<br><br>When combined with the option of sample size re-estimation and/or adjustment of the stopping boundaries during the trial, these designs are sometimes referred to as 'adaptive group sequential'. |
| Sample size re-estimation | Sample size reassessment / recalculation / adaptations | Adjustment of the sample size, in either a blinded or unblinded fashion. This is typically done to help ensure the trial achieves the desired power.<br><br>Sample size re-estimation is sometimes combined with other trial adaptations, such as group sequential testing. |

| Multi-arm multi-stage (MAMS) | Pick-the-winner / drop-the-loser / treatment selection | Compares multiple treatments, doses, durations or combinations to a common control, allowing for early stopping of treatment arms for efficacy or futility. In addition, this can include interim treatment selection of the most promising treatment(s) for further investigation in the subsequent stages.<br><br>Variations of MAMS designs include 'drop-the-loser' trials, where inferior treatment arms are dropped from the trial (with the control group typically retained). 'Seamless' trials (such as phase II/III trials) are also closely related, where the selection and confirmatory phases of development are combined into one trial. 'Platform' trials can also be viewed as an extension of MAMS designs, by allowing new treatment arms to be added to the trial over time. |
|---|---|---|
| Response-adaptive randomisation (RAR) | Outcome-adaptive randomisation | Shifts allocation ratios towards more promising or informative treatment(s) based on the accumulating patient response data. RAR is sometimes combined with options for early stopping (e.g., for futility or efficacy). |
| Adaptive enrichment | Population enrichment / patient enrichment / adaptive population enrichment | Allows for selection of a target population during the trial, typically using pre-defined patient subpopulations based on biomarker information. For example, one class of adaptive enrichment designs allows the eligibility criteria to be adaptively updated during the trial based on the estimated treatment-biomarker interactions, restricting entry in the subsequent trial stage(s) to patients most likely to benefit from the new treatment. |

*Table 1: Definitions of the different types of adaptive designs considered in this paper.*

# 2. Estimation bias in adaptive designs

The issue with estimation after an adaptive trial is that traditional maximum likelihood (ML) estimators tend to be biased either because of some selection that took place following an interim analysis (see Bauer et al.[30] for a detailed explanation of why selection results in bias) or other mechanisms utilised in an adaptive design, such as early stopping, which might affect the sampling distribution of the estimator (these depend on the nature of the design). For this reason, the usual ML estimator (MLE) is sometimes referred to as the 'naive' estimator for the trial.

Before considering the issue of estimation further it is worth clarifying what we mean by a biased estimator. The FDA guidance on adaptive designs[23] defines bias as "a systematic tendency for the estimate of treatment effect to deviate from its true value", and states that "It is important that clinical trials produce sufficiently reliable treatment effect estimates to facilitate an evaluation of benefit-risk and to appropriately label new drugs, enabling the practice of evidence-based medicine". It is clear that (all else being equal) it is desirable to obtain estimators of treatment effects that are unbiased in order to make reliable conclusions about the effects of study treatments.

While it is relatively easy to define statistical bias, different definitions of an unbiased estimator are relevant in our context. To introduce these, let us denote the population parameter of interest, the treatment effect, by $\theta$ and an estimator thereof by $\widehat{\theta}$.

*Mean-unbiased estimators*

An estimator $\widehat{\theta}$ is called *mean-unbiased* if its expected value is the same as the true value of the parameter of interest, that is, $E(\widehat{\theta}) = \theta$. This is the most commonly used definition of unbiasedness.

*Median-unbiased estimators*

An estimator $\widehat{\theta}$ is called *median-unbiased* if $P(\widehat{\theta} < \theta) = P(\widehat{\theta} > \theta)$, that is if the probability of overestimation is the same as the probability of underestimation. Note that for symmetric sampling distributions of $\widehat{\theta}$, a median-unbiased estimator is also mean-unbiased.

*Conditionally and unconditionally unbiased estimators*

A further distinction of unbiasedness in estimators refers to whether they are *conditionally* or *unconditionally* unbiased. In our context, an estimator is unconditionally unbiased (also known as marginally unbiased), if it is unbiased when averaged across all possible realisations of an adaptive trial. In contrast, an estimator is conditionally unbiased if it is unbiased only conditional on the occurrence of a subset of trial realisations. For example, one might be interested in an estimator only conditional on a particular arm being selected at an interim analysis; as such, the focus becomes on a conditional unbiased estimator. We discuss the issue of conditional versus unconditional bias further throughout the rest of this paper.

*Bias and mean squared error*

When considering estimation after an adaptive trial one is often faced with the following conundrum: precise estimators can potentially be biased while unbiased estimators tend to be less precise, which reflects the classical bias-variance trade-off. This means that in many instances the mean squared error (MSE), a measure of precision defined as $E((\widehat{\theta} - \theta)^2)$, of a biased estimator is often smaller than the MSE of an unbiased estimator. This makes it challenging to find the 'best' estimator that fits a particular trial design.

*The focus of our review*

Since bias as defined above is an expectation (or probability) taken over possible realisations of data, it is inherently a frequentist concept. However, we can still evaluate the frequentist bias of a Bayesian point estimator, such as the posterior mean. Similarly, it is possible to use frequentist point estimators in the analysis of Bayesian trial designs (i.e., where the adaptations and decision rules are driven by Bayesian analyses). Hence we consider both Bayesian point estimators and Bayesian adaptive designs in this paper. However, we restrict our attention to phase II and III trial designs, since phase III trial results have a direct impact on health policy and the adoption of new treatments for wider public use, while phase II trial results directly influence whether further research of a treatment in phase III is required.

## 3. Methodological review: search strategy and paper selection

We conducted a database search of Scopus on 13th July 2022 of all available papers up to that date. We used a "title, abstract, keywords" search, with the pre-defined search terms for different categories of estimator given in Table 2.

| **Category of estimator** | **Search terms** |
|---|---|
| Unbiased | ((((unbiased OR ((conditionally OR mean OR median OR "uniformly minimum variance" OR "uniformly minimum variance conditionally") AND unbiased)) AND estim*) OR "median adjusted" OR umvcue OR umvue) AND ( "adaptive design" OR "adaptive trial" OR "adaptive clinical trial" OR "group-sequential")) |
| Bias-reduced | ((("bias-reduced" OR "bias-adjusted" OR "conditional moment" OR "empirical Bayes" OR shrinkage) AND estimat*) OR ((corrected OR adjusted OR conditional) AND (mle OR "maximum likelihood"))) AND ("adaptive design" OR "adaptive trial" OR "adaptive clinical trial" OR "group-sequential") |
| Resampling-based methods | (((bootstrap OR nonparametric OR non-parametric OR resampling OR "distribution free" OR distribution-free OR jackknife OR "Monte Carlo") AND estim*) AND ("adaptive design" OR "adaptive trial" OR "adaptive clinical trial" OR "group-sequential")) |

*Table 2: Search strategy used for the initial database search of Scopus.*

We excluded papers not involving adaptive clinical trial designs, as well as those that did not propose or evaluate unbiased or bias-reduced estimators. Two authors (DSR and PP) separately reviewed the papers, extracted data and checked each other's work for accuracy.

Our search strategy retrieved a total of 257 papers, of which 148 were excluded immediately as they were not relevant based on the title and abstract. We then looked for additional relevant papers citing or cited by these remaining 109, which added 59 papers. After completing a full text review of these papers, a total of 145 were deemed relevant, and information about the trial contexts, advantages, limitations, code availability and case studies was extracted for qualitative synthesis (Figure 1). Full results giving a summary of each paper extracted for qualitative analysis are found in the supporting materials.

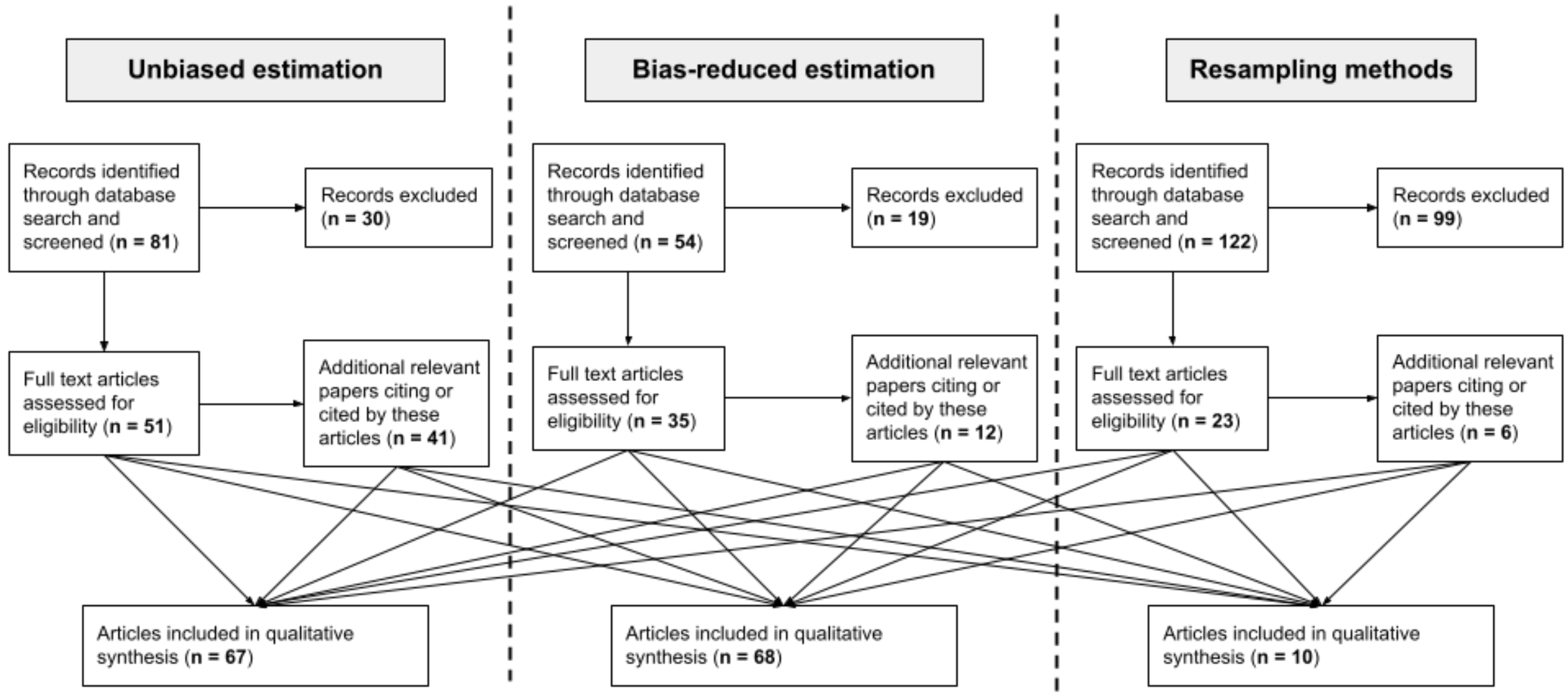

*Figure 1: Diagram showing the flow of information through the different phases of the systematic review. Note that Bayesian and empirical Bayes methods are included in the 'bias-reduced' category.*

## 4. Comparisons of estimators, trial examples and software

In this section, we summarise the results of our systematic review in Table 3 by classifying them according to the broad class of adaptive design used. For each design class, we give references to the relevant literature for the different types of estimators that have been proposed. Where the literature focuses on designs that allow for multiple adaptive features within a single trial, we have used our judgement to place them into a class of adaptive designs, and refer the reader to Sections 5 and 6 for further clarifying details. In Table 3, we also give a summary of general pros and cons of the different estimators and the comparisons between them as given in the literature. Finally, we point out examples of the use of different estimators on trial data, as well as where software/code is available. Note that this list of software/code is by no means exhaustive, but only includes those that were specifically mentioned in the reviewed methodological literature or those of which the authors were already aware.

The systematic review identified mean-unbiased estimators, median unbiased estimators, resampling-based estimators and bias-reduced estimators. The bias-reduced estimators include those that estimate bias and subtract it from the MLE and those that use shrinkage methods. Detailed descriptions and explanations of the different types of point estimators referenced in Table 3 are given in the following two sections, with Section 5 focusing on unbiased estimation (both mean-unbiased and median-unbiased) and Section 6 focusing on bias-reduced estimation (including Bayesian approaches and resampling-based methods).

| Design | Method(s) | Pros and cons | Trial examples and software/code |
|---|---|---|---|
| **Group sequential** | **Mean-unbiased estimation**<br><br>*Chang et al. (1989)*[31], *Kim (1989)*[32], *Emerson and Fleming (1990)*[33], *Emerson and Kittelson (1997)*[34], *Liu and Hall (1999)*[35], *Jung and Kim (2004)*[36], *Liu et al. (2006, 2007)*[37,38], *Porcher and Desseaux (2012)*[39], *Zhao et al. (2015)*[40].<br><br>Secondary parameters: *Gorfine (2001)*[41], *Liu and Hall* [42], *Liu et al. (2005)*[43], *Kunz and Kieser (2012)*[44].<br><br>Diagnostic tests: *Shu et al. (2008)*[45], *Pepe et al. (2009)*[46]. | The UMVUE has zero bias, but tends to have a higher MSE than the naive or bias-adjusted estimators.<br><br>Computation can be complex and extensive if the number of looks is relatively large ($\geq 4$), but otherwise can be simpler than for bias-adjusted estimators.<br><br>UMVUE can be conditionally biased. | Beta-Blocker Heart Attack Trial, see Gorfine (2001)[41].<br><br>Phase II trial in patients with adenocarcinoma, see Kunz and Kieser (2012)[44].<br><br>GI06-101 trial in hepatobiliary cancer, see Zhao et al. (2015)[40].<br><br>*Software/code*<br>OptGS R package |
| | **Median-unbiased estimation**<br><br>*Kim (1988, 1989)*[32,47], *Emerson and Fleming (1990)*[33], *Todd et al. (1996)*[48], *Troendle and Yu (1999)*[49], *Hall and Liu (2002)*[50], *Koyama and Chen (2007)*[51], *Wittes (2012)*[52], *Porcher and Desseaux (2012)*[39], *Shimura et al. (2017)*[53].<br><br>Secondary parameters: *Hall and Yakir (2003)*[54].<br><br>Adaptive group sequential trials: *Wassmer (2006)*[55], *Brannath et al. (2009)*[56], *Gao et al. (2013, 2014)*[57,58], *Levin et al. (2014)*[59], *Mehta et al. (2019)*[60], *Nelson et al. (2022)*[61]. | MUE reduces bias compared to the naive estimator, but can have an increased MSE. Bias-adjusted estimators can have a lower MSE as well.<br><br>Calculation of the MUE can be complicated, and results can depend on the ordering of the sample space.<br><br>MUEs can be derived for adaptive group sequential designs, unlike for other estimation methods. | Trials for acute bronchitis reported in Wassmer et al. (2001)[62].<br><br>Multicenter Automatic Defibrillator Implantation Trial (MADIT), see Hall and Liu (2002)[50], Hall and Yakir (2003)[54].<br><br>Non-small cell lung cancer trial, see Wassmer (2006)[55].<br><br>Randomized Aldactone Evaluation Study (RALES), see Wittes (2012)[52].<br><br>Trial reported in Troger et al. |

| | | | (2013)[63].<br><br>Clinical Evaluation of Pertuzumab and Trastuzumab (CLEOPATRA) trial, see Shimura et al. (2017)[53].<br><br>*Software/code*<br>[ADDPLAN]<br>[Levin et al. (2014)]<br><br>Note: also implemented in software such as SAS, East, as well as R packages such as rpact, RCTdesign, AGSDest and OptGS. See https://panda.shef.ac.uk/techniques/group-sequential-design-gsd/categories/27 |
|---|---|---|---|
| | **Resampling**<br><br>Parametric<br>*Pinheiro and DeMets (1997)*[64], *Wang and Leung (1997)*[65], *Leung and Wang (1998)*[66], *Cheng and Shen (2013)*[67], *Magnusson and Turnbull (2013)*[17].<br><br>Non-parametric<br>*Leblanc and Crowley (1999)*[68]. | Essentially the same procedure can be used under different stopping rules and different study designs.<br><br>Bootstrap algorithms can be computationally intensive.<br><br>Bias is substantially reduced, with reasonable MSE.<br><br>Non-parametric approaches are robust to model misspecification. | Trial for nasopharyngeal cancer, see Leblanc and Crowley (1999)[68]. |
| | **Bias-reduced**<br><br>*Whitehead (1986)*[69], *Chang et al. (1989)*[31], *Tan and Xiong (1996)*[70], *Todd et al. (1996)*[48], *Li and DeMets (1999)*[71], *Fan et al. (2004)*[72], *Liu et al. (2004)*[73], *Guo and Liu (2005)*[74], *Porcher and Desseaux (2012)*[39], *Shimura et al. (2018)*[75], *Li (2011)*[76]<br><br>Adaptive group sequential trials: *Levin et al. (2014)*[59].<br><br>Conditional: *Troendle and Yu (1999)*[49], *Coburger and Wassmer (2001)*[77] | The MSE of the bias-adjusted MLE is typically lower than that of the UMVUE, particularly for small sample sizes.<br><br>Shrinkage-type estimators (which also have a Bayesian interpretation) can reduce both the conditional bias and MSE further | Three different phase II studies, see Tan and Xiong (1996)[70].<br><br>Trial of immunosuppression for bone marrow transplantation, see Whitehead (1986)[78].<br><br>Two cardiovascular trials (MERIT-HF and COPERNICUS), see Fan et al. (2004)[72].<br><br>Trial in familial adenomatous polyposis, see Liu et al. (2004)[73]. |
| | Secondary parameters: *Whitehead (1986)*[78], *Liu et al. (2004, 2005)*[43,73], *Yu et al. (2016)*[79]<br><br>Health economic outcomes: *Flight (2020)*[80]. | | Phase II trial in endometrial cancer, see Shimura et al. (2018)[75].<br><br>GUSTO trial for myocardial infarctions, see Marschner and Schou[81].<br><br>*Software/code*<br>RCTdesign R package<br>OptGS R package<br>Levin et al. (2014). |
| | **Bayesian**<br><br>*Hughes and Pocock (1988)*[82], *Pocock and Hughes (1989)*[83] | Useful for producing shrinkage of unexpectedly large/imprecise observed treatment effects that arise in trials that stop early.<br><br>Bias reduction depends on the specification of the prior distribution. | None |
| **Sample size re-estimation** | **Mean-unbiased estimation**<br><br>Unconditional*Liu et al. (2008)*[84], *Liu et al. (2012)*[85], *Kunzmann and Kieser (2017)*[86].<br><br>Conditional (UMVCUE)<br>*Kunzmann and Kieser (2017)*[86], *Broberg and Miller (2017)*[87]. | Estimators have zero bias, either unconditionally or conditionally. However, the MSE tends to be greater than for the naive estimator and bias-reduced estimators.<br><br>Does not guarantee compatibility with the test decision (see Kunzmann and Kieser[86]).<br><br>Estimators can have an explicit representation making computation easy. | Schizophrenia trial, see Broberg and Miller (2017)[87]. |
| | **Median-unbiased estimation**<br><br>*Lawrence and Hung (2003)*[88], *Liu et al. (2008)*[84], *Wang et al. (2010)*[89], *Liu et al. (2012)*[85], *Kunzmann and Kieser (2017)*[86], *Nhacolo and Brannath (2018)* [90].<br><br>Conditional perspective<br>*Broberg and Miller (2017)*[87]. | MUE tends to have small mean bias, and can also have smaller MSE than the naive estimator.<br><br>Does not guarantee compatibility with the test decision.<br><br>MUE can be calculated for flexible adaptation rules that are not completely | Coronary artery disease trial, see Wang et al. (2010)[89].<br><br>Trial on reperfusion therapy for acute myocardial infarction, see Brannath et al. (2006)[95].<br><br>*Software/code*<br>R packages such as rpact and |
| | Flexible sample size adaptations<br>*Bauer et al. (2001)*[91], *Liu and Chi (2001)*[92], *Brannath et al. (2002)*[93], *Lawrence and Hung (2003)*[88], *Proschan et al. (2003)*[94], *Brannath et al. (2006)*[95]. | pre-specified in advance, unlike for other estimation methods. | adpss. |
| | **Bias-reduced**<br><br>*Denne (2000)*[96], *Coburger and Wassmer (2003)*[97], *Cheng and Shen (2004)*[98], *Shen and Cheng (2007)*[99], *Liu et al. (2008)*[84], *Tremmel (2010)*[100], *Broberg and Miller (2017)*[87]. | Proposed bias-reduced estimates are nearly unbiased with practical sample sizes, with similar variance to the naive estimator.<br><br>Numerical problems can occur when calculating adjusted estimators, and observations close to the critical boundaries can lead to unreasonably extreme adjusted estimators. | Type II Coronary Intervention Study, see Denne (2000)[96].<br><br>Colon cancer trial, see Shen and Cheng (2007)[99].<br><br>Trial in chronic lymphocytic leukemia, see Tremmel (2010)[100].<br><br>Schizophrenia trial, see Broberg and Miller (2017)[87]. |
| | **Bayesian**<br><br>*Kunzmann and Kieser (2017)*[86], *Grayling and Mander (2022)*[101] | Guarantees compatibility with the test decision.<br><br>Reduces MSE of MLE except for very small or very large values of the success probability.<br><br>Reduces absolute bias compared with MLE except for small values of the success probability, where there can be a substantial positive bias.<br><br>Can reduce MSE substantially compared to the UMVUE for certain response rates. | None |
| **Multi-arm multi-stage designs (with treatment selection)** | **Mean-unbiased estimation**<br><br>Two-stage designs<br>*Cohen and Sackrowitz (1989)*[102], *Tappin (1992)*[103], *Bowden and Glimm (2008, 2014)*[104,105], *Pepe et al. (2009)*[46], *Koopmeiners et al. (2012)*[106], *Robertson et al. (2015, 2016)*[107,108], *Robertson and Glimm (2018)*[109]<br><br>Multi-stage designs | UMVCUEs are conditionally unbiased. Compared to the MLE, the conditional MSE tends to be lower, but unconditionally the MSE can substantially increase<br><br>UMVCUEs in the literature tend to have a closed-form expression, allowing for easy computation | Trial for the treatment of anxiety disorder, see Kimani et al. (2013)[111] and Robertson et al. (2016)[108]<br><br>INHANCE study, see Robertson and Glimm (2018)[109]<br><br>ADVENT trial, see Stallard and |

| | *Bowden and Glimm (2014)*[105], *Stallard and Kimani (2018)*[110]<br><br>Seamless phase II/III trials<br>*Kimani et al. (2013)*[111], *Robertson et al. (2016)*[108] | In some settings, the UMVCUE can have comparable MSE to bias-adjusted estimators | Kimani (2018)[110]<br><br>PROVE trial[112] - implements approach of Stallard and Kimani (2018)[110]<br><br>*Software/code*<br>Bowden and Glimm (2014) |
|---|---|---|---|
| | **Resampling**<br><br>*Pickard and Chang (2014)*[113], *Whitehead et al. (2020)*[114] | Provides a reasonable balance between bias and MSE across several scenarios<br><br>Approach can be applied to endpoints coming from a variety of distributions (including normal and binomial)<br><br>Approaches are robust to model mis-specification | *Software/code*<br>Whitehead et al. (2020) |
| | **Bias-reduced**<br><br>Two-stage designs<br>*Coad (1994)*[115], *Shen (2001)*[116], *Stallard et al. (2008)*[117], *Pepe et al. (2009)*[46], *Luo et al. (2010, 2012)*[118,119], *Bebu et al. (2010, 2013)*[120,121], *Koopmeiners et al. (2012)*[106], *Brückner et al. (2017)*[122]<br><br>Multi-stage designs<br>*Coad (1994)*[115], *Stallard and Todd (2005)*[123], *Bowden and Glimm (2014)*[105]<br><br>Seamless phase II/III trials<br>*Kimani et al. (2013)*[111] | Bias-adjusted MLE can have relatively low MSE and acceptably small bias in some scenarios<br><br>Shrinkage methods can be the most effective in reducing the MSE<br><br>Bias-reduced estimators can run into computational/convergence problems.<br><br>Estimators can overcorrect for bias | Phase II study in colorectal cancer, see Luo et al. (2010)[118]<br><br>*Software/code*<br>Luo et al. (2010)[118]<br><br>FOCUS trial in advanced colorectal cancer, see Brückner et al. (2017)[122]<br><br>Phase III trial in Alzheimer's, see Stallard and Todd (2005)[123]<br><br>*Software/code*<br>Bowden and Glimm (2014)<br><br>*Software/code*<br>Kimani et al. (2013) |
| | **Bayesian**<br><br>Two-stage designs<br>*Carreras and Brannath (2013)*[124], *Bowden et al. (2014)*[125], *Brückner et al. (2017)*[122] | Shrinkage estimators can perform favourably compared with the MLE in terms of bias and MSE | FOCUS trial in advanced colorectal cancer, see Brückner et al. (2017)[122] |
| | Multi-stage designs<br>*Bunouf and Lecoutre (2008)*[126] | Approaches can run into practical and theoretical complications, necessitating further modifications such as estimating a between-arm heterogeneity parameter | |
| **Response-adaptive randomisation** | **Mean-unbiased estimation**<br><br>*Bowden and Trippa (2017)*[127] | Mean-unbiased estimators can have a large MSE and can be very computationally intensive to calculate. | Glioblastoma trial with multiple treatments |
| | **Bias-reduced**<br><br>*Coad (1994)*[128]*, Morgan (2003)*[129]*, Marschner (2021)*[130] | The conditional MLE can be very effective at eliminating the conditional bias that is present in the unconditional MLE, but this comes at the cost of a loss of efficiency except for more extreme designs.<br><br>The penalised MLE exhibits very little conditional bias and is not subject to substantial efficiency loss (compared to the unconditional MLE) when the realised design is close to its average. | None |
| **Adaptive enrichment designs** | **Mean-unbiased estimation**<br><br>Two-stage designs<br>*Kimani et al. (2015, 2018, 2020)*[131–133]*, Kunzmann et al. (2017)*[134]*, Di Stefano et al. (2022)*[135] | UMVCUE is unbiased, but tends to have a higher MSE than the MLE<br><br>UMVCUE can also have a larger MSE than shrinkage/bias-adjusted estimators, but compensates by the corresponding eradication of bias<br><br>UMVCUEs are easily computable, as they have a closed form expression | MILLY phase II study in asthma, see Kunzmann et al. (2017)[134]<br><br>*Software/code*<br>Kimani et al. (2020) |
| | **Bias-reduced**<br><br>Two-stage designs<br>*Kunzmann et al. (2017)*[134]**,** *Kimani et al. (2018)*[132]*, Di Stefano et al. (2022)*[135] | Can have lower MSE than UMVCUE that is comparable to the MSE of the MLE, but has residual bias | MILLY phase II study in asthma, see Kunzmann et al. (2017)[134] |
| | **Resampling**<br><br>*Magnusson and Turnbull (2013)*[17]*,* | Bootstrap estimator can have a higher | MILLY phase II study in |

| | *Kunzmann et al. (2017)*[134], *Simon and Simon (2018)*[136] | bias than the MLE in the two-stage setting of Kunzmann et al. (2017)[134], but Simon and Simon (2018)[136] found it was very effective at correcting for bias in their multimarker setting<br><br>Procedures can be computationally intensive | asthma, see Kunzmann et al. (2017)[134] |
|---|---|---|---|
| | **Bayesian**<br><br>Two-stage designs<br>*Kunzmann et al. (2017)*[134], *Kimani et al. (2018)*[132], *Di Stefano et al. (2022)*[135] | Estimators exhibit a higher bias than the MLE in many situations, with varying MSE properties | MILLY phase II study in asthma, see Kunzmann et al. (2017)[134] |

*Table 3: Summary of unbiased and bias-reduced estimators, trial examples and software/code for different classes of adaptive designs. MLE = Maximum Likelihood Estimator; MSE = Mean Squared Error; MUE = Median Unbiased Estimator; UMVCUE = Uniformly Minimum Variance Conditionally Unbiased Estimator; UMVUE = Uniformly Minimum Variance Unbiased Estimator.*

Looking at the results as a whole, we can see that the literature on unbiased and bias-adjusted estimation for adaptive designs has grown rapidly in recent years, reflecting the increased use of adaptive designs in practice. However, only a few of these methodological papers present real-life trial examples (such as re-analyses of trials that have already been completed), and even fewer provide software/code. In terms of the different classes of adaptive designs, group sequential designs have received the most attention by far, and some of the estimation methods are implemented in widely-available statistical software such as R or SAS. This reflects the relatively long history of group sequential designs and their more widespread use in practice. In contrast, we found very little literature on estimation for trials using response-adaptive randomisation. With the increasing use of such trials (see e.g. the ISPY-2[137] and BATTLE[138] trials), this may be an important gap in the literature to fill.

Another potential gap is that some of the methods are restricted to two-stage designs (especially for adaptive enrichment trials), and so a natural extension would be to develop these methods for multi-stage trials. As well, most of the methodology has focused on binary and normally-distributed outcomes, with comparatively few proposals tailored for trials with time to event outcomes. It is also unclear to what extent existing methodology can apply to longitudinal outcomes (which may be particularly common for interim analyses) or to cluster randomised trials.

In terms of different types of estimators, generally speaking there is a bias-variance tradeoff (see the start of Section 6) unbiased estimators may pay the 'price' for complete unbiasedness by having a high variance. Conversely, bias-reduced estimators may have a lower variance than unbiased estimators, but may "overcorrect" and thereby introduce bias in the opposite direction. Another general observation is that mean-unbiased estimators in the literature require pre-specified boundaries and decision rules, and so are not applicable to flexible adaptive designs with arbitrary changes. Finally, resampling-based methods have received comparatively little attention as opposed to other types of bias-reduced estimation. For example, as far as we could tell, there has not been work looking explicitly at their use for sample size re-estimation designs. Given that resampling-based methods can be applied even to complex trial designs (see e.g., Whitehead et al.[114]), their use could potentially provide one solution to the problem of estimation for trials which combine different types of adaptations together and hence will not fit neatly into the classes of adaptive designs used above.

# 5. Unbiased estimation

## 5.1 Mean-unbiased estimation

In order to achieve exact mean-unbiasedness, typically it is necessary to find an estimator whose distribution is independent of the trial adaptations. One key setting where this is possible is in classical group sequential trials. Given that there is a pre-specified number of observations at the time of the first interim analysis, then the sample mean (which corresponds to the MLE) at the end of the first stage is (unconditionally) unbiased as no adaptations to the trial have occurred. However, this estimator is also clearly inefficient for estimating the overall treatment effect, as it does not use any information that may arise from later stages of the trial.

Another setting where unbiased estimation is possible are multi-stage trials with ranking and selection from a subset of candidates, such as candidate treatments or patient subgroups. The sample mean (MLE) for the selected candidate(s) calculated using only the final stage data will be conditionally unbiased, conditional on the ranking and selection that has taken place in the previous stages. However, again this estimator is clearly inefficient as it ignores all the information about the selected candidate(s) from the previous stages of the trial. Moreover, unlike the first stage estimator described in the previous paragraph, it cannot be used to estimate the effect of a treatment or subpopulation that was dropped at an interim analysis.

*Minimum variance unbiased estimation*
One way to obtain a more efficient estimator that uses more of the information from the trial, while still maintaining exact mean-unbiasedness, is to apply the Rao-Blackwell theorem. This theorem implies that if U is an unbiased estimator of the unknown parameter of interest $\theta$, and T is a sufficient statistic, then the estimator $\hat{U} = E(U|T)$ is unbiased and $\text{var}(\hat{U}) \leq \text{var}(U)$. In certain cases, this 'Rao-Blackwellisation' technique allows the derivation of the unbiased estimator that has the smallest possible variance, which is known as the uniformly minimum variance unbiased estimator (UMVUE). The Lehmann-Scheffé theorem[139] states that if $\mathscr{T}$ is a sufficient and complete statistic, and U is an unbiased estimator, then the Rao-Blackwellised estimator $\hat{U} = E(U|\mathcal{T})$ is the unique UMVUE.

The derivation of UMVUEs (from an unconditional perspective) has been a focus in the literature on group sequential trials, with work by Chang et al.[31] and Jung and Kim[36] for binary response data (see also Porcher and Desseaux[39] and Zhao et al.[40] for Simon's two-stage designs), and Kim[32], Emerson and Fleming[33] and Emerson and Kittelson[34] for normally-distributed endpoints. Liu and Hall[35] proved that the sufficient statistic for the latter setting is in fact not complete, but that the Rao-Blackwellised estimator is still UMVUE among all 'truncation-adaptable' unbiased estimators – i.e., where inference following stopping early does not require knowledge of the future analyses. For simplicity, in what follows we do not make this technical distinction when describing UMVUEs. Similar, more general theoretical results can be found in Liu et al.[37] for group sequential tests for distributions in a one-parameter exponential family and Liu et al.[38] for multivariate normal group sequential tests. See also Liu et al.[140] and Liu and Pledger[141] for further theoretical results for general two-stage adaptive designs.

Methodology has also been developed to calculate the UMVUE for secondary parameters or endpoints in group sequential trials. Liu and Hall[42] derived the UMVUE in the context of correlated Brownian motions, while Gorfine[41] considered normally-distributed endpoints where the secondary parameter is the mean in a subgroup of subjects. Similarly, Liu et al.[43] derived the UMVUE for a secondary probability, such as the rate of toxicity, in group sequential trials with binary endpoints. Kunz and Kieser[44] also derived UMVUEs for secondary endpoints in two-stage trials with binary endpoints. Further settings include estimating the sensitivity and specificity of a diagnostic test using a group sequential design, where UMVUEs have been derived by Shu et al.[45] and Pepe et al.[46].

Moving away from classical group sequential designs, Liu et al.[85] derived the UMVUE for a two-stage adaptive design (with sample size adjustment based on conditional power). Kunzmann and Kieser[86] derived the UMVUE general two-stage adaptive designs with binary endpoints, where the second stage sample size is determined by a function of the response rate observed in stage one. Meanwhile, Liu et al.[84] proposed Rao-Blackwellised unbiased estimators for both primary and secondary endpoints in the context of a two-stage adaptive trial with sample size re-estimation. Bowden and Trippa[127] showed how to calculate a Rao-Blackwellised unbiased estimator for trials using RAR with binary endpoints.

In some trial contexts, it may be more appropriate to find unbiased estimators *conditional* on the adaptations that have taken place. For example, in multi-stage trials with treatment selection, typically there would be greater interest in estimating the properties of the better-performing treatments that are selected at the end of the trial, rather than those treatments that are dropped for futility. In addition, it may not be possible to find a complete statistic for the parameter of interest without additionally conditioning on the selection rule used (see e.g., Cohen and Sackrowitz[102]). For both these reasons, in settings such as multi-stage trials with ranking and selection, there has been a focus on deriving uniformly minimum variance *conditionally* unbiased estimators (UMVCUEs, also known as CUMVUEs), where the conditioning is on the triggered adaptations.

One of the first papers to take this conditional perspective and calculate the UMVCUE was Cohen and Sackrowitz[102], in the context of a two-stage drop-the-loser trial with treatment selection and normally-distributed endpoints (see Tappin[103] for the setting with binary endpoints), where only the best-performing treatment is taken forward to the final stage. This was subsequently extended by Bowden and Glimm[104] to allow the calculation of the UMVCUE for treatments that were not the best-performing, Bowden and Glimm[105] for multi-stage drop-the-loser trials (this was only a Rao-Blackwellised estimator) and Robertson and Glimm[109] for unknown variances. Meanwhile Koopmeiners et al.[106] derived the UMVCUE for a two-stage trial evaluating continuous biomarkers. A similar line of work focused on deriving UMVCUEs for binary data, with Pepe et al.[46] looking at two-stage designs testing the sensitivity of a dichotomous biomarker (this UMVCUE can also be applied for Simon's two stage designs, see Porcher and Desseaux[39], which was extended by Robertson et al.[107]).

For seamless phase II/III trials with a normally-distributed endpoint, Kimani et al.[111] showed how to calculate a Rao-Blackwellised conditionally unbiased estimator, which was subsequently extended by Robertson et al.[108]. More recently, Stallard and Kimani[110] derived the UMVCUE for MAMS trials (again with normally-distributed endpoints) with treatment selection and early stopping for futility, conditional on any pre-specified selection or stopping rule.

A recent setting where UMVCUEs have been derived is two-stage adaptive enrichment designs, where biomarkers are used to select a patient population to investigate in the second stage. Kimani et al.[131] and Kunzmann et al.[134] derived the UMVCUE for such designs with normally-distributed endpoints, which was extended by Kimani et al.[132] to allow the biomarker cut-off to be determined by the first stage data, and Kimani et al.[133] and Di Stefano et al.[135] for time-to-event endpoints. Finally, in the context of two-stage adaptive designs where the second-stage sample size depends on the result of the first stage through a pre-specified function, Kunzmann and Kieser[86] and Broberg and Miller[87] considered UMVCUEs for normally-distributed and binary endpoints, respectively.

## 5.2 Median-unbiased estimation

While exact mean-unbiasedness is a common criterion to aim for, it is not always feasible or desirable to use mean-unbiased estimators. The main potential disadvantage is an increase in the MSE when compared with the usual MLE, due to the bias-variance trade-off. As well, the calculation of Rao-Blackwellised estimators may become difficult or infeasible for more flexible trial designs, where adaptations are not fully pre-specified in advance. For both these reasons, median-unbiased estimators (MUEs) have been proposed, which can have a lower MSE than mean-unbiased estimators in some settings, and can be derived for a wider class of adaptive trial designs.

Consider testing a hypothesis about some parameter of interest θ. One common approach to calculating MUEs is as follows:

1. Define an ordering of the trial design space with respect to evidence against $H_0$. For example, in group sequential trials the predominant option is stage-wise ordering: this depends on the boundary crossed, the stage at which stopping occurs and the value of the standardised test statistic (in decreasing order of priority).

2. Using this ordering, define a *p*-value function P(θ) which gives the probability that, at the stage the trial stopped, more extreme evidence against $H_0$ could have been observed.

3. At the point the trial stops, find the MUE $\widehat{\theta}_{MU}$, which satisfies $P(\widehat{\theta}_{MU}) = 0.5$.

The last step is essentially finding a 50% confidence bound for θ, or equivalently the midpoint of a symmetric 100(1 - α)% confidence interval (constructed using the *p*-value function, which takes into account the previous adaptations or stopping rules).

A number of papers have derived MUEs for group sequential trials, including Kim[32,47], Emerson and Fleming[33], Todd et al.[48] and Troendle and Yu[49]. Hall and Liu[50] provided MUEs that account for overrunning, while Hall and Yakir[54] derived MUEs for secondary parameters. Note also early work by Woodroofe[142] showed how to calculate a MUE but in the context of sequential probability ratio tests. Meanwhile, MUEs for Simon's two-stage design were derived by Koyama and Chen[51].

For adaptive group sequential trials Wassmer[55] and Nelson et al.[61] showed how to calculate MUEs for normally-distributed endpoints, while Brannath et al.[56] derived a MUE for survival endpoints. These results were subsequently extended by Gao et al.[58], who presented general theory for point and interval estimation (see also Mehta et al.[60]). Gao et al.[57] also derived a MUE for an adaptive group sequential design that tests for noninferiority followed by testing for superiority.

Moving away from group sequential trials, Wang et al.[89] and Liu et al.[85] focused on two-stage adaptive designs with sample size adjustments based on conditional power considerations, and showed how to calculate MUEs in these settings. Lawrence and Hung[88] proposed a MUE for the setting where the maximum information of a trial is adjusted on the basis of unblinded data (see also Liu et al.[84] for an improved version of this estimator). More generally, MUEs for two-stage trials with an adaptive sample size based on a pre-specified function of the interim results are given by Kunzmann and Kieser[86] and Nhacolo and Brannath[90]. Meanwhile, Li et al.[143] derived MUEs in the context of '2-in-1' adaptive designs, which allow for a seamless expansion of a phase II trial into a phase III trial or an expansion of biomarker positive populations, where the endpoints for the interim and final analyses can be different.

Like for mean-unbiased estimation, the conditional perspective has also been used for median-unbiased estimation. The previously mentioned work by Hall and Yakir[54] for group sequential trials is an early example of this conditional perspective, while the approach of Koopmeiners et al.[106] can also be used to calculate a conditional MUE for two-stage group sequential designs. In the specific context of two-stage group sequential designs with survival endpoints, Shimura et al.[53] showed how to calculate conditional MUEs, for the cases where the study stops or does not stop early for either efficacy or futility. Meanwhile, Broberg and Miller[87] considered the general setting of a two-stage sample size adjustable design based on interim data, and derived the MUE conditional on continuing to the second stage.

MUEs can be derived for flexible adaptive designs, which allow arbitrary changes to the trial at each stage based on the interim results and/or external information. Bauer et al.[91] first described the construction of MUEs for two-stage designs based on combination tests and conditional error functions; see also Liu and Chi[92], Proschan et al.[94] and Brannath et al.[95]. Finally, Brannath et al.[93] gave a general method to calculate MUEs when using recursive combination tests for flexible multi-stage adaptive trials.

# 6. Bias-reduced estimation

Whilst using an unbiased estimator whenever there is one available for a particular adaptive design may seem like a straightforward choice, this may not always be the best option, due to the inherent bias-variance tradeoff: reducing the bias of an estimator can come at the cost of increasing its variance (thus decreasing its precision) substantially. For example, estimation procedures that involve conditioning on a statistic that is informative about the parameter of interest (e.g., the stopping time) address the bias, but discard the information that is contained in that statistic and hence lead to a loss of efficiency. The bias-variance tradeoff can lead to unbiased treatment effect estimates being accompanied by large standard errors and wide confidence intervals, which may be of little use practically. Therefore, reducing the bias but not completely eliminating it sometimes proves to be a better solution than aiming for complete unbiasedness.

## 6.1 Analytical and iterative procedures

Cox recognised as early as 1952[144] that in sequential test procedures the option to stop early introduces bias to the MLE. They proposed a modified MLE which can be used for repeated significance tests[145] but not more general group sequential tests. Whitehead[69] focused on the sequential probability ratio test[146] and the triangular test[147], and derived an analytic expression to quantify the bias and proposed a bias-adjusted estimator $\tilde{\theta}$ obtainable by subtracting the bias from the MLE:

$$\tilde{\theta} = \hat{\theta} - bias(\tilde{\theta})$$

Evaluating the bias at $\tilde{\theta}$ leads to an iterative procedure, and this method can be used for various types of endpoints including continuous, binary, and time-to-event. Guo and Liu[74] argued that a simpler single-iteration version could be used, where the estimated bias of the MLE is subtracted rather than the bias at the unknown $\tilde{\theta}$:

$$\tilde{\theta}_s = \hat{\theta} - bias(\hat{\theta})$$

They suggested that this approach could be extended from single-arm phase II trials to more general estimation problems involving designs with early stopping. Note that these proposals aim to reduce the unconditional bias of the MLE, see below for the conditional perspective.

Chang et al.[31] and applied Whitehead's idea to group sequential designs with binary response data, while Tan and Xiong[70] proposed using the bias-adjusted MLE to estimate response rates in trials using fully sequential and group sequential conditional probability ratio tests. Li and DeMets[71] derived an exact analytical expression for the bias of the MLE for a group sequentially monitored Brownian motion process with a linear drift, and used it to construct a bias-adjusted estimator based on the arguments of Whitehead[69]. Todd et al.[48] studied the bias-adjusted MLE further in simulations of group sequential designs with a normally-distributed endpoint using the O'Brien-Fleming or triangular test. Denne[96] postulated that the same bias-adjusted MLE could also be used to estimate the treatment effect following a two-stage sample size re-estimation design on the basis of conditional power, which was subsequently investigated by Liu et al.[84] (who also considered a bias-adjusted MLE for secondary endpoints). Levin et al.[59] found that the bias-adjusted MLE could be applied to a wider class of adaptive group sequential designs with pre-defined rules.

Whitehead[78] extended the idea of the bias-adjusted MLE to the estimation of secondary endpoints in sequential trials using two different approaches: conditional on the primary endpoint (in which case the MLE is unbiased) and unconditional. Liu et al.[43] further extended Whitehead's approach to estimate secondary probabilities after termination of a sequential phase II trial with response rate as the primary endpoint. The bias-adjusted MLE has also been proposed for a group sequential strategy to monitor toxicity incidents in clinical trials by Yu et al.[79], for which they noted the bias of the usual MLE was "considerably large". Meanwhile, Flight[80] showed how to use the bias-adjusted MLE for health economic outcomes following a group sequential design.

For trials with treatment selection, early work by Coad[115] in the setting with two treatments and either a two or three-stage trial (where only one treatment is selected to be carried forward to the final stage), showed how to derive a bias-adjusted MLE. In the context of adaptive enrichment designs with time-to-event endpoints, Di Stefano[135] derived both single and multiple iteration versions of a bias-adjusted MLE. Finally, for response-adaptive trials, Coad[128] and Morgan[129] showed how to approximate the bias of the MLE, which then allows the construction of a bias-adjusted estimator following the idea of Whitehead[69].

*Conditional perspective*

Troendle and Yu[49] developed methods for estimating treatment differences in group sequential designs with normally-distributed endpoints that condition on the stopping time $S$ of the trial, thereby reducing the conditional bias, i.e., the discrepancy between the conditional expectation of the estimator and the parameter value:

$$conditional\,bias(\theta) = E(\widehat{\theta}|S) - \theta$$

They used a similar iterative approach as Whitehead[69] to compute the estimator. Coburger and Wassmer[97] took this one step further, by conditioning both on stopping time as well as the sample sizes of the stages up to that point; whereas Troendle and Yu had assumed equal sample sizes for all stages. They also considered adaptive group sequential designs, where the unconditional bias is no longer computable, but the conditional bias can be approximated. Tremmel[100] extended these considerations to stratified binomial and time-to-event endpoints. Coburger and Wassmer[77] suggested using a similar conditional bias-adjusted estimator not only when a trial stops but also during the course of an adaptive sample size re-estimation design, to prevent an overestimation of the sample sizes for subsequent stages.

Fan et al.[72] showed that the conditional bias-reduced estimator for group sequential designs as proposed by Troendle and Yu[49] is equivalent to a conditional MLE as well as to a conditional moment estimator. They also proposed two modified estimators which are hybrids of the conditional and unconditional MLE that have smaller variance and MSE. The relationship between the conditional MLE and the conditional bias-reduced estimator was also pointed out by Liu et al.[73], who additionally derived the conditional MLE for secondary endpoints in a group sequential design and adaptive group sequential design. More general theoretical results for sequential trials focusing on the conditional MLE can be found in Molenberghs et al.[148] and Milanzi et al.[149,150].

Marschner[130] recently extended this notion of conditional MLEs to more general adaptive designs, as part of a unifying formulation of adaptive designs and a general approach to their analysis. More specifically, the unconditional likelihood can be expressed as the product of the design likelihood (i.e., the information contained in the realised design) and the conditional likelihood (conditional on the realised design). Marschner also proposed a penalised MLE, which weights the design likelihood and can vary between the two extremes of the unconditional and conditional MLEs. Examples of the conditional and penalised MLEs were given in the context of a RAR design.

Shimura et al.[75] modified Troendle and Yu's conditional bias-reduced estimator for two-stage group sequential trials with normal endpoints by combining it with the shrinkage idea of Thompson[151]: in the calculation of the bias adjustment term, they replaced the MLE with a weighted average of the MLE at the interim analysis and a 'prior' estimate of the effect size, $\widehat{\theta}_* = c\widehat{\theta} + (1 - c)\theta_0$, with $\theta_0$the effect size assumed in the initial sample size calculation and shrinkage weights $c$ determined as a function of the MLE as proposed by Thompson[151]. Although the authors presented their idea in a frequentist framework, it can also be interpreted as Bayesian, with $\theta_0$the prior and $\widehat{\theta}_*$ the posterior expectation. Meanwhile, Li[76] proposed using a weighted sum of the MLE and sample proportion for two-stage group sequential designs with binary responses, where the weight is based on estimates of the MSEs of these estimators. Pepe et al.[46] and Koopmeiners et al.[106] considered conditional bias-reduced estimation in the context of group sequential diagnostic biomarker studies with a single interim analysis for futility, conditional on not stopping early.

Cheng and Shen[98] derived an easily computable bias-adjusted estimator of the treatment effect in a 'self-designing' trial (where the sample size is sequentially determined to achieve a target power) with a normally-distributed endpoint[152]. They further proposed a modified estimator, involving a shrinkage factor, to reduce bias when block sizes are small. For 'self-designing' trials with censored time-to-event endpoints and group sequential sample size re-estimation[153], Shen and Cheng[99] developed bias-adjusted estimators of the hazard ratio. Broberg and Miller[87] considered two-stage sample size adjustable designs with normally-distributed endpoints and derived a conditional bias-adjusted MLE. Meanwhile, in the context of '2-in-1' adaptive trial designs, Li et al.[143] derived a conditionally bias-adjusted estimator.

Moving on to trial designs with treatment selection, when multiple treatments are compared to a control and the most promising treatment is selected at the first interim analysis (with or without further early stopping opportunities at subsequent interim analyses), Stallard and Todd[123] developed a bias-adjusted estimator for the treatment effect. Kimani et al.[111] adapted this estimator for the difference between the selected and control treatments in seamless phase II/III designs with normally-distributed endpoints. Unlike Stallard and Todd's estimator[123], which aimed to be approximately unbiased conditional on the selected treatment, this new estimator aimed to be approximately unbiased conditional on both the selected treatment and continuation to the second stage. In the context of selecting the treatment dose with the highest response rate (using a normal approximation), Shen[116] also derived a stepwise bias overcorrection. When there are two treatments being compared with a control, Stallard et al.[117] proposed a family of approximately conditionally unbiased designs.

Brückner et al.[122] extended Stallard and Todd's bias-reduced estimator to two-stage multi-arm designs involving a common control and treatment selection at the interim analysis with time-to-event endpoints. Despite deriving analytical expressions for the selection bias of the MLE in the selected and stopped treatment arms, respectively, they experienced computational difficulties and convergence problems. They only defined the resulting bias-reduced estimator for the interim analysis, and proposed using a two-stage estimator for the final analysis, which is a weighted average of the bias-reduced estimator at the interim analysis and the MLE.

Luo et al.[118] used the method of conditional moments to derive a bias-adjusted estimator of the response rate following a two-stage drop-the-loser design with a binary endpoint. Adopting a stochastic process framework (as opposed to the conventional random variable viewpoint) for clinical trials with adaptive elements, Luo et al.[119] proposed approximating treatment effects based on a general estimating equation to match the first conditional moment. This generic approach can be used for a variety of adaptive designs with endpoints following any probability distribution. They illustrated it for two-stage treatment selection designs comparing multiple treatments against a control.

Kunzmann et al.[134] applied the conditional moment estimator of Luo et al.[118] to two-stage adaptive enrichment designs with a normally-distributed endpoint and binary biomarker with a pre-specified cut-off. Additionally, they studied two hybrid estimators, one between the UMVCUE and the conditional moment estimator, and the other one combining the UMVCUE and MLE. Kimani et al.[132] studied single- and multi-iteration bias-adjusted estimators for the treatment effect in the selected subpopulation of a two-stage adaptive threshold enrichment design allowing the cut-off of a predictive biomarker based on first stage data.

Finally, Bebu et al.[120] devised a bias-corrected conditional MLE for the mean of the selected treatment in two-stage designs and normally distributed endpoints. They focused on the case of two experimental treatments, but this was extended by Bebu et al.[121] to designs involving the selection of more than one treatment arm, selection rules based on both efficacy and safety, including a control arm, adjusting for covariates, and binomial endpoints. Bowden and Glimm[105] further extended the method to multi-stage drop-the-loser designs which must always proceed to the final stage (i.e., no early stopping rules).

## 6.2 Bayesian and empirical Bayes approaches

Several Bayesian approaches to bias-reduced estimation in adaptive designs have been developed, typically for trials using frequentist frameworks. These differ from fully Bayesian adaptive designs in that they only use the accumulating data, rather than the posterior, to make decisions about adaptations (e.g., stopping or selecting arms) and rely primarily on classical frequentist inference such as hypothesis testing, but often require additional assumptions or information, such as the specification of a prior.

Hughes and Pocock[82] and Pocock and Hughes[83] proposed a Bayesian shrinkage method for estimating treatment effects in trials stopped at an interim analysis, whereby a prior quantifying the plausibility of different treatment effect sizes is specified at the outset of the trial and combined with the trial data using Bayes' rule, producing a posterior estimate of the treatment effect that is shrunk towards the median of the prior distribution. They acknowledged that this will be sensitive to the choice of prior and warned against using an overoptimistic prior as this would lead to little or no shrinkage, saying that their method "works best in the hands of prior specifiers who are realists by nature"[82].

Hwang[154] proposed using Lindley's[155] shrinkage estimator for estimating the mean of the best treatment in single-stage multi-arm designs with 4 or more study treatments and a normally-distributed endpoint. This was extended by Carreras and Brannath[124] to multi-arm two-stage drop-the-losers designs. Their two-stage version is a weighted mean of Lindley's estimator for the first stage and the MLE of the selected treatment for the second stage. It shrinks the MLE towards the mean across all arms, but only works with a minimum of 4 study treatment arms. For designs with 2 or 3 study treatments, they replaced Lindley's estimator with the best linear unbiased predictor of a random effects model for the first stage estimator. This method also allows for bias-reduced estimation of the second best treatment mean.

Of note, Lindley's estimator can be viewed as an empirical Bayes estimator of the posterior mean of the posterior distribution of a specific treatment mean given its MLE. This fact was exploited by Bowden et al.[125], who adapted Carreras and Brannath's method and replaced their two-stage estimator with a single shrinkage equation for both stages. This, however, adds both practical and theoretical complications, necessitating further modifications such as estimating a between-arm heterogeneity parameter. The authors suggest this approach could be applied to other MAMS designs.

Along the lines of Hwang's approach[154], Brückner et al.[122] derived two shrinkage estimators for two-stage multi-arm designs with a common control, treatment selection at the interim analysis, and time-to-event endpoints. They both shrink the MLE of the selected treatment towards the overall log-hazard ratio. The authors also defined a two-stage estimator similar to that of Carreras and Brannath[124], as well as a bias-corrected Kaplan-Meier estimator for the selected treatment.

Bunouf and Lecoutre[126] derived two Bayesian estimators for multi-stage designs with binary endpoints where after every stage a decision is made whether to stop or continue recruiting: one estimator is the posterior mean based on a 'design-corrected' prior, the other the posterior mode based on a similarly corrected uniform prior, and for the latter they also provided an easy-to-use approximation for two-stage designs. Meanwhile, Kunzmann et al.[134] developed two empirical Bayes estimators for two-stage adaptive enrichment designs with a single pre-specified subgroup and pre-specified decision rule based on a binary biomarker and a normally-distributed endpoint with known variance. Kimani et al.[132] extended this work to adaptive threshold enrichment designs where the biomarker cut-off is determined based on first stage data.

Finally, Kunzmann and Kieser[86] developed an estimator for two-stage single-arm trials with a binary endpoint with sample size adjustment and early stopping for futility or efficacy. This estimator minimises the expected MSE subject to compatibility with the test decision and monotonicity conditions, and can be viewed as a constrained posterior mean estimate. Similarly, in the same trial context Grayling and Mander[101] proposed an optimal estimator that minimises a weighted sum of the expected MSE and bias.

## 6.3 Resampling-based methods

The methods for unbiased and biased-reduced estimation considered in the previous subsections require explicit formulae based on the population response model as well as the trial design and adaptations. For example, UMVUEs and UMVCUEs can often be given as closed-form expressions, while for MUEs a *p*-value function is often specified. Meanwhile, bias-reduced estimators tend to have explicit expressions for an estimate of the bias, leading to equations that can be solved numerically.

An alternative class of methods is based on *resampling* procedures, where the trial data is resampled or generated via a parametric bootstrap a large number of times, and the resulting trial replicates can then be used to give empirical estimates of the bias. One advantage of these methods is that essentially the same procedure can be used under a variety of different stopping rules and trial designs, including complex designs as detailed below.

In the group sequential setting, Wang and Leung[65] showed how to use a parametric bootstrap to calculate a bias-corrected estimate for the mean of normally-distributed endpoints with either known or unknown variance. This methodology was generalised by Leung and Wang[66], who proposed a generic stochastic approximation approach based on a parametric bootstrap, which can be used for non-independent and identically distributed (iid) data and a large class of trial designs, including sequential ones. Cheng and Shen[67] proposed a variant of the Wang and Leung[65] procedure for complex group sequential monitoring rules that assess the predicted power and expected loss at each interim analysis. Magnusson and Turnbull[17] proposed a double bootstrap procedure for group sequential enrichment designs incorporating subgroup selection. Meanwhile, Pinheiro and DeMets[64] described a simulation approach to estimate the bias of the MLE and then proposed constructing a bias-reduced estimator for the treatment difference, although this is equivalent to Whitehead's bias-adjusted MLE[69].

Kunzmann et al.[134] carried out further work in the context of adaptive enrichment trials. In the two-stage trial setting with a binary biomarker and normally-distributed endpoints, they considered a number of alternatives to the MLE, including a parametric bootstrap procedure. More generally, Simon and Simon[136] considered multi-stage enrichment designs that develop model-based predictive classifiers based on multiple markers, and showed how to use a parametric bootstrap method for bias correction for generic response distributions. Finally, in trials with treatment selection, Pickard and Chang[113] considered parametric bootstrap corrections for a two-stage drop-the-losers design, for normally- and binomially-distributed endpoints.

Thus far, all the methods presented above are fully parametric, i.e., assuming that the trial outcome data comes from a probability distribution with a fixed set of parameters. An alternative approach is to use *non-parametric* bootstrap procedures to correct for bias. An early example of this approach was given by Leblanc and Crowley[68], who proposed a non-parametric bootstrap in the context of group sequential designs with censored survival data using log rank testing. Subsequently, Rosenkranz[156] described using non-parametric bootstrap to correct the bias of treatment effect estimates following selection (e.g., selecting the treatment with the maximum effect). More recently, Whitehead et al.[114] proposed an estimation approach based on the method of Rao-Blackwellisation (i.e., targeting unbiased estimation), but used a non-parametric bootstrap procedure to recreate replicate observations in a complicated group sequential trial setting comparing multiple treatments with a control.

# 7. Discussion

Motivated by recent FDA guidance on estimation after adaptive designs, this paper provides a review of the potential methodological solutions to biased estimation that exist in the literature. We found that there is a growing body of work proposing and evaluating a range of unbiased and bias-adjusted estimators for a wide variety of adaptive trial designs. Our ambition is that this paper, combined with the annotated bibliography given in the Supplementary Information, provides an easily-accessible and comprehensive resource for researchers in this rapidly evolving area. However, we note that statistical software and code to easily calculate adjusted estimators is relatively sparse, which is an obstacle to the uptake of methods in practice (see Grayling and Wheeler[157] and part II of this paper series). It also remains the case that for more complex or novel adaptive designs, adjusted estimators may not exist.

From a methodological perspective, there remain important open questions regarding estimation after adaptive trials, as already mentioned at the end of Section 6. More generally, an estimator for a parameter of interest should ideally have all of the following properties:

1. Adequately reflects the adaptive design used;
2. No or small bias;
3. Low MSE (reflecting a favourable bias-variance trade-off);
4. Is easily computable;
5. A procedure for calculating an associated confidence interval that:

a. Has the correct coverage; and
b. Is consistent / compatible with the hypothesis test decisions (including early stopping)

Constructing an overarching framework for estimation after adaptive designs that has all these properties would be very useful, although challenging. Some initial steps in this direction have been taken by Kunzmann and Kieser[86,158] in the context of adaptive two-stage single-arm trials with a binary endpoint. Even if such a framework is not feasible more generally, the issue of constructing confidence intervals remains an important question that has so far received less attention in the literature than point estimation.

The property that an estimator should 'adequately reflect the adaptive design used' also has implications when considering the Bayesian analysis of adaptive designs, as pointed out by an anonymous reviewer. The likelihood function will typically be insensitive to the adaptive design and sampling scheme (see e.g., Section 4.1 of Marschner 2021[130]). Thus, while the frequentist interpretation will take account of the adaptive design, the pure Bayesian approach following the likelihood principle will not. In practice this would mean that a pure Bayesian analysis would ignore the adaptive nature of the design and the issues discussed in this paper would disappear. However, this (arguably) is intuitively unappealing and provides a rationale for evaluating Bayesian-motivated approaches from a frequentist perspective.

In part II of this paper series, we explore the practical considerations surrounding the use of unbiased and bias-reduced estimators for adaptive designs. There, we review the use of such estimators in current practice and illustrate their application to a real trial design. We also provide a set of guidelines for best practice, considering their use in adaptive trials from the design stage through to the final reporting of results.

## Acknowledgements

This research was supported by the NIHR Cambridge Biomedical Research Centre (BRC-1215-20014). This report is independent research supported by the National Institute for Health Research (Prof Jaki's Senior Research Fellowship, NIHR-SRF-2015-08-001). The views expressed in this publication are those of the authors and not necessarily those of the NHS, the National Institute for Health Research or the Department of Health and Social Care (DHCS). T Jaki and DS Robertson received funding from the UK Medical Research Council (MC_UU_00002/14). DS Robertson also received funding from the Biometrika Trust. B Choodari-Oskooei was supported by the MRC grant (MC_UU_00004_09 and MC_UU_12023_29). The Centre for Trials Research receives infrastructure funding from Health and Care Research Wales and Cancer Research UK.

## Data Availability Statement

All of the data that support the findings of this study are available within the paper and supplementary information. For the purpose of open access, the author has applied a Creative Commons Attribution (CC BY) licence to any Author Accepted Manuscript version arising.